\documentclass[prl,twocolumn,floatfix,showpacs,superscriptaddress]{revtex4}
\usepackage{amsfonts}
\usepackage{amsmath}
\usepackage{amssymb}
\usepackage{graphicx}
\usepackage{natbib}
\usepackage[colorlinks=true]{hyperref}

\begin{document}
\title{The phonon softening due to melting of the ferromagnetic order in elemental iron }
\author{Qiang Han}
\affiliation{Department of Physics \& Astronomy, Rutgers University, Piscataway, NJ 08854-8019, USA}
\author{Turan Birol}
\affiliation{Department of chemical engineering and materials science,University of Minnesota}
\author{Kristjan Haule}
\affiliation{Department of Physics \& Astronomy, Rutgers University, Piscataway, NJ 08854-8019, USA}

\date{\today}

\begin{abstract}
We study the fundamental question of the lattice dynamics of a metallic ferromagnet in the regime where the static long range magnetic order is replaced by  the fluctuating local moments embedded in a metallic host.
We use the \textit{ab initio} Density Functional Theory(DFT)+embedded Dynamical Mean-Field Theory(eDMFT) functional approach
to address the dynamic stability of iron polymorphs and the phonon softening with increased temperature.
We show that the non-harmonic and inhomogeneous phonon softening measured in iron is a result of the melting of the long range ferromagnetic order, and is unrelated to the first order structural transition from the BCC to the FCC phase, as is usually assumed. We predict that the BCC structure is dynamically stable at all temperatures at normal pressure, and is only thermodynamically unstable between the BCC-$\alpha$ and the BCC-$\delta$ phase of iron.
\end{abstract}

\maketitle

The theoretical description of the interplay between structural, magnetic and electronic degrees of freedom in transition metals at finite temperatures is a central problem of condensed matter physics. The elemental iron is the archetypical system to study the coupling of the ferromagnetism and electronic degrees of freedom with the crystal structure, and its importance in both the geophysics at high temperatures and high pressures, and metallurgy at normal pressure but finite temperature, has made iron one of the most thoroughly studied materials. Its magnetic and mechanical properties undergo major changes through a series of structural phase transitions, but the clear understanding of the feedback effect of magnetism on the structural stability has been elusive.

Elemental iron crystalizes in four different polymorphs, among them are two body-centered cubic (BCC) phases and a face-centered-cubic (FCC) phase,  all realized at normal pressure. The BCC-$\alpha$ phase is stable below $1185\,$K, the FCC-$\gamma$ phase follows and is stable up to $1670\,$K, where it is transformed to the BCC-$\delta$ phase, which is stable up to the melting point around $1811\,$K.  The $\alpha$ phase is ferromagnetic (FM) below the Curie temperature $T_c=$1043K.

Many theoretical methodologies to describe energetics of magnetic materials, and stability of different allotropes, have been developed over the past few decades. 
The conventional Density Functional Theory (DFT), in its GGA approximation, predicts quite well the magnetic properties of the FM BCC structure with correct moment and good bulk modulus, and quite accurate phonon spectra. However GGA severely underestimates the stability of the competing non-magnetic FCC phase, which is around 300meV higher in energy than FM BCC phase, and is predicted to be dynamically unstable in its nonmagnetic phase, with many imaginary phonon branches~\cite{GCLDA}. Similarly, the high-temperature BCC $\delta$-phase is dynamically and thermodynamically unstable within this standard approach.

To simulate the interplay between the lattice dynamics and presence of magnetic moments at finite temperature in metallic environment, several approaches have been developed, which broadly fall into three categories: i) considering static magnetic configurations within DFT, but disordered in real space~\cite{SQS,PMphonon1,Katsnelson,Kormann_2014}, ii) supplementing DFT energetics by some information obtained by an auxiliary Heisenberg model, which is exactly solved by the quantum Monte Carlo method~\cite{Heisenberg_M,Kormann_2014}, iii) dynamic many body approaches, such as the DMFT, which simulate the dynamics on a single site exactly, but neglect the exchange-correlation energy between different iron sites.~\cite{review,Anisimov,Lichtenstein}

To mimic the presence of local moments within static DFT, K\"ormann et al.~\cite{PMphonon1} developed a methodology for calculating phonon frequencies at very high-temperature in real-space large unit cell disordered simulation, by employing space averaging within constrained spin-DFT. This approach is closely related to special quasi-random structures methodology~\cite{SQS0}. They showed that in such real-space disordered state, both the BCC and FCC structures become dynamically stable, and that phonons are considerably softer in this high-temperature state than in ferromagnetic state. Ikeda et.al~\cite{SQS} used the same method to study the pressure dependence of phonons spectra. K\"ormann et.al.~\cite{Kormann_2014} later extended this method to treat  the paramagnetic phase as a function of temperature using auxiliary Heisenberg model simulation.
It was shown that such approach can describe reasonably well the temperature dependent phonon softening measured in experiment~\cite{Phonon_softeningexp,AnharmonicsPhonon}. 

A related method, based on large unit cell DFT calculations, was used in Ref.~\onlinecite{Katsnelson} and \onlinecite{anharmonic} to study pressure and temperature dependence of phonon spectra. In this method, the disorder in atomic positions is coming from thermal vibrations of the lattice, rather than from the disorder in spin orientations, hence it includes anharmonic effects due to phonon-phonon interaction. It was noticed in Ref.~\onlinecite{anharmonic} that non-magnetic disordered state simulations predict both BCC and FCC structures to be dynamically unstable. However, when simulation is performed in the fictitious long-range antiferromagnetic state, the results are in good agreement with experiment, even when structural disorder is switched off. The inclusion of lattice dynamical effects improved the agreement with experiment slightly, but it is clearly not the main force in stabilizing the high temperature phases of iron. It was thus shown that the presence of magnetic moments, and their role in lattice energetics, is far more important than the thermal disorder in lattice position.

While the above described studies based on DFT static simulations, but with inclusion of the real-space spin disorder are broadly consistent with experimental measurements, their validity relies on the ergodicity of the quantum metallic system. The local fluctuating magnetic moments are disordered in time rather than space, as their Bragg peaks do not show extra broadening beyond standard thermal disorder, hence proper treatment of fluctuating moments has to be dynamic.  With the advent of the dynamical mean field theory (DMFT) and its combination with DFT, the nature of electrons which are partially itinerant, forming metallic bands in iron, and partly localized, giving rise to Curie-Weiss susceptibility, could finally be simulated from \textit{ab-initio}~\cite{PRL_Kotliar}. 

The energetics of BCC to FCC transition in iron has been addressed by DMFT method in Ref.~\onlinecite{PRL_Leonov}, and was later extended to study lattice dynamics in the paramagnetic state of BCC and FCC structure~\cite{PMphonon2}, but the lattice dynamics of the ferromagnetic state has not been addressed before. Moreover, the authors of Ref.~\onlinecite{PMphononnature} recently ascribe the previously observed phonon softening  at the $N$-point in the Brillouin zone ~\cite{Phonon_softeningexp,AnharmonicsPhonon} to the correlations effects changing with the temperature in the paramagnetic state of the system~\cite{PMphononnature}. They predicted that the paramagnetic BCC structure becomes dynamically unstable between 1.2-1.4 times Curie temperature (close to the $\alpha$ to $\gamma$ transition in iron) and gets progressively more unstable in most branches as temperature is increased, so that  the BCC-$\delta$ phase would require large phonon-phonon interaction to be dynamically stabilized. Consequently they conclude that the $\alpha$-$\gamma$ transition in iron occurs due to this phonon-softening at the N-point.

So far the phonon calculations by DMFT~\cite{PMphonon2,PMphononnature,Savrasov} were performed only in the paramagnetic state. On the other hand, the DFT based methods~\cite{SQS,anharmonic} always require some sort of static order, hence the effects of melting the long range magnetic order with temperature, and the impact of partially ordered and disordered local magnetic moments on phonons was not properly addressed before, and is the focus of this study. 
Moreover, previous DFT+DMFT calculations for iron~\cite {PMphonon2,PMphononnature} were using non-stationary implementation of DFT+DMFT total energy expression, which is based on the intermediate downfolded auxiliary Hubbard model, and hence the force does not appear as a derivative of a stationary functional.  The stationary implementation of DFT+embedded DMFT functional has recently been achieved~\cite{stationary} and its analytic derivative, which gives rise to the force, was derived in Ref.~\onlinecite{force_Haule}, hence the force contains the effects of electronic and magnetic entropy, missing in the previous DMFT approaches. The resulting phonon dynamics, which includes the effects of finite temperature electronic and magnetic entropic effects, is hence more trustworthy in the high temperature paramagnetic phases than previous reports.

The physical picture emerging from this state of the art computational technique is very different from previous DMFT reports: i) the first order phase transition from alpha to gamma phase is unrelated to observed phonon softening in iron. ii) The experimentally observed softening of phonons and their non-harmonic change is a consequence of the melting of the long range ferromagnetic order, and once the paramagnetic state is reached, the change of the phonons with temperature is reasonably well explained by the quasi-harmonic approximation.  iii) The BCC state remains dynamically stable at all temperatures even though the FCC state is thermodynamically the stable phase between $\alpha$ and $\delta$ phase.
Consequently the phonon-phonon interaction is not needed to make the high temperature BCC-$\delta$ phase dynamically stable.

In this letter we use the stationary version of DFT+embedded DMFT method~\cite{review, DFTDMFT,website} 
in which the forces are derivatives of the stationary free energy functional with respect to ion displacement.~\cite{force_Haule}
The continuous time quantum Monte Carlo in its rotationally invariant form is used as the impurity solver~\cite{ctqmc,wernerctqmc}.  The screened value of the Coulomb interaction 
is determined by the constrained LDA method resulting in $U=5.5~eV$ and Hund's exchange interaction $ J=0.84~eV$~\cite{Anisimov_cLDA}, and we used the nominal double counting, which was shown to be very close to exact double-counting~\cite{PRL_DC}. The DFT part is based on Wien2k package\cite{wien2k} and we use LDA functional which, when combined with DMFT, predicts better crystal structures. This is because in LDA functional both the electronic bandwith and equilibrium lattice constants consistently show signatures of overbinding, and can both be corrected by adding dynamic correlations, while in GGA the bandwidth shows similar overbinding tendency, while lattice constants many times shows underbinding tendency, hence they are harder to simultaneously correct by higher order theory.
The phonon spectrum is calculated using direct approach as implemented in the phonopy package~\cite{Togo20151}.

\begin{figure}[!t]
\centering{
\includegraphics[width=0.99\linewidth,clip=]{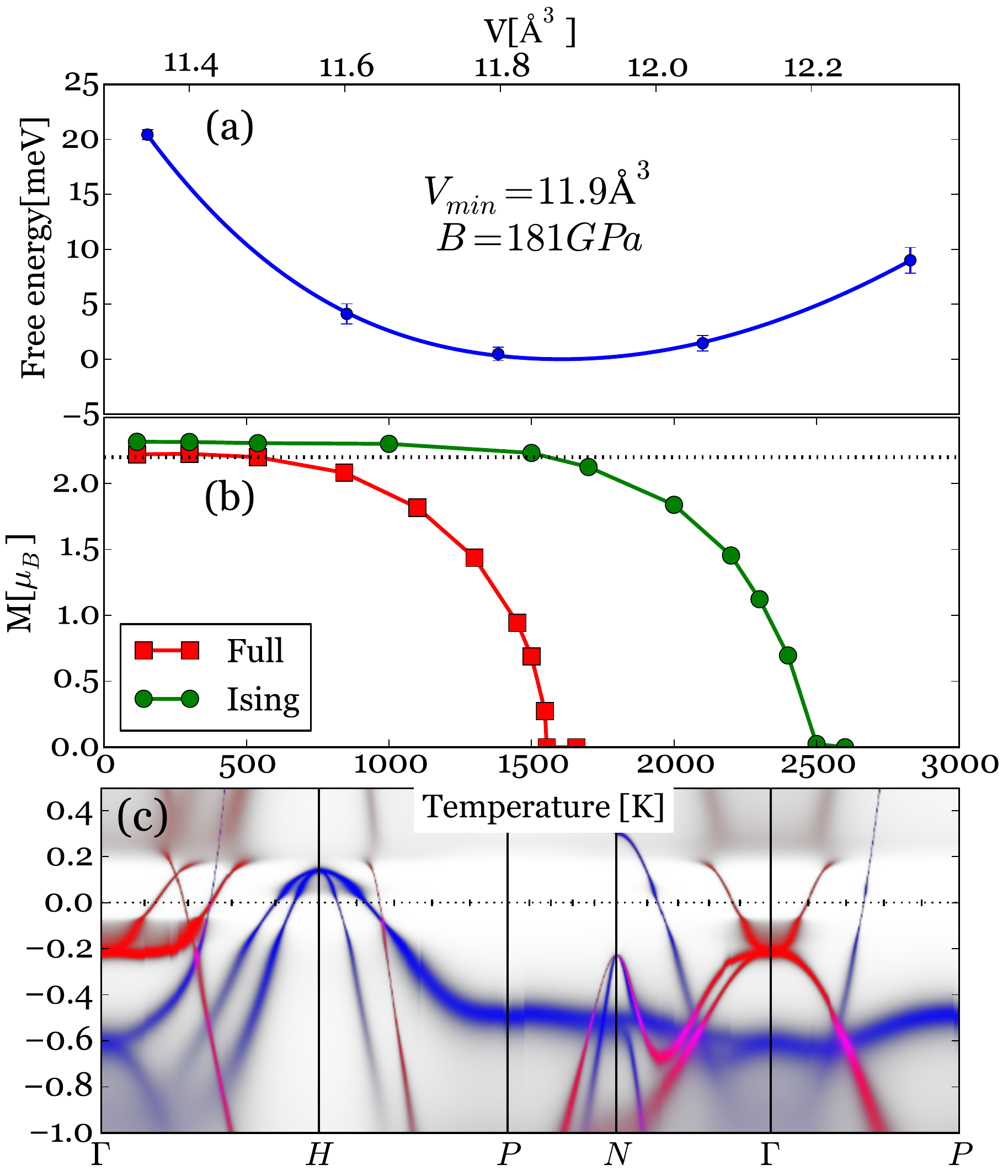}
}
\caption{(a) The electronic free energy per atom versus $V$ of cubic unit cell.
(b) The temperature dependence of the ordered ferromagnetic moment of bcc iron using both the density-density (``Ising") and the rotationally-invariant (``Full")  Coulomb interaction form. (c) The single-particle spectral function of the BCC-$\alpha$ phase at $300\,$K (the majority and minority spectra are plotted in blue and red color, respectively).
}
\label{fig1}
\end{figure}
In Fig.~\ref{fig1}(a) we show the free energy versus volume of BCC unit cell at room temperature, which gives the equilibrium volume $11.9~\textrm{\AA}^3$  and bulk modulus  $181~\textrm{GPa}$, which are in good agreement with experimental values of $11.69~\textrm{\AA}^3$ \cite{Basinski459} and $172~\textrm{GPa}$ \cite{GCLDA}. Fig.~\ref{fig1}(b) shows the magnetization versus temperature curve, which follows the mean field type of behavior, and gives almost exact magnetic moment 2.2~$\mu_B$. The transition temperature ($T^{Full}_c=$1550K) in this direct calculation is overestimated, as expected for a method which treats spatial correlations on a mean-field level, consequently the phase with a short range order is typically predicted to have stable long range order. We also show the same magnetization curve for the case when the Coulomb interaction is approximated with the density-density terms only (Ising approximation) to demonstrate that such approach, which was previously used in Refs. ~\onlinecite{PRL_Leonov,PMphonon2,PMphononnature}  leads to much higher transition temperature and somewhat larger magnetic moment. This effect was also noticed in Ref.~\onlinecite{mag3,mag4} using Hirsch-Fye quantum Monte Carlo method, but was neglected in previous studies of lattice dynamics. In Fig.~\ref{fig1}(c) we also show the electronic spectral function at $300\,$K, which is in very good agreement with ARPES measurement of Ref.~\onlinecite{FeARPES} (see Supplementary material for more detail, which includes additional Refs.~\onlinecite{LWF,Baym,HFF,ironsc1,ironsc2,ironsc3,ironsc4,TMO1,TMO2,TMO3,TMO4,TMO5,lan1,lan2,lan3,act1,act2,iri1,iri2,iri3,kat,FLL,IPS,UJpara,cLDAwien2k,cLDA1,cLDA2,cRPA1,cRPA2,cRPA3,smode}), in contrast to earlier DMFT calculations based on approximate impurity solvers~\cite{old_ARPES}.
We note that similar magnetization curve for iron was shown in Ref.~\onlinecite{PRL_Kotliar} using reduced temperature and reduced moment, but here we show that the same interaction parameters lead to very precise absolute value of the magnetic moment and correct equilibrium lattice constant, as well as the correct renormalization of the electronic band structure, as measured by ARPES.

\begin{figure}[!th]
\centering{
\includegraphics[width=0.9\linewidth,clip=]{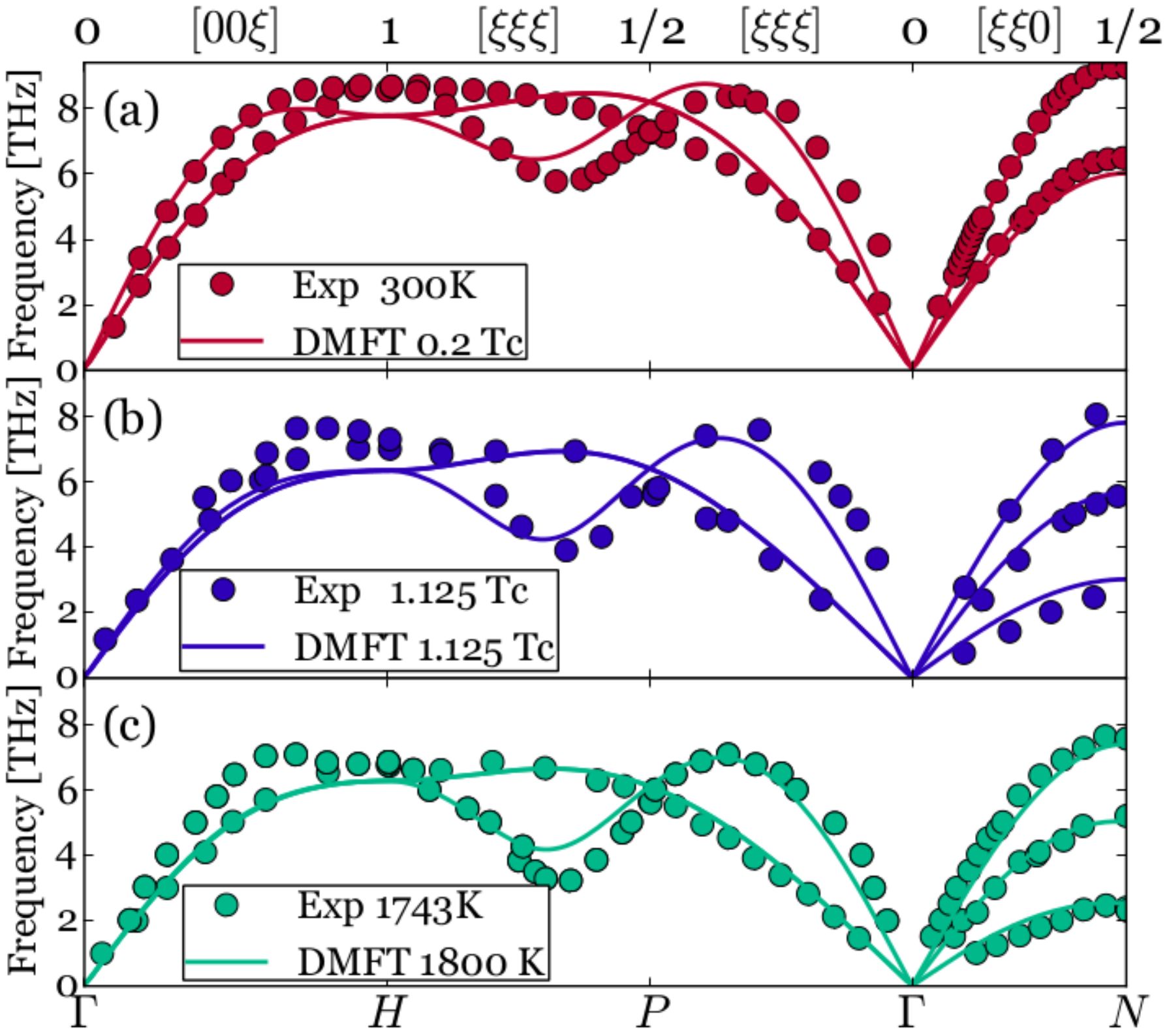}
}
\caption{Phonon spectrum at  low temperature ($T = 300 K$), in the paramagnetic BCC-$\alpha$ phase ($T= 1.125 T_c$) and in the paramagnetic BCC-$\delta$ phase ($T=1800\,$K) evaluated at experimental lattice constants. The dots correspond to the experimental data from Refs.~\onlinecite{phononT300K}, ~\onlinecite{Phonon_softeningexp}  and ~\onlinecite{PMphononPRB} 
at $300\,$K, $1.125\,$T$_c$ and $T=1743\,$K, respectively. 
}
\label{fig2}
\end{figure}
In Fig.~\ref{fig2} we show the phonon spectra calculated in the three BCC phases of iron, in the ferromagnetic state at room temperature, in the paramagnetic $\alpha$ state slightly above the Curie temperature (1.125$\,T_c$), and in the BCC-$\delta$ phase at high temperature, and we compare it to the measured spectra from Refs.~\onlinecite{phononT300K,Phonon_softeningexp,PMphononPRB}. They are compared at the same scaled temperature $T/T_c$ in FM state, as $T_c$ is overestimated in our calculation, while in the paramagnetic state ($T\sim 1800K$) we 
use absolute temperature, 
because electronic structure above $T_c$ depends primarily
on the lattice constant (which is taken from the experiment). 
We notice reasonable agreement between the theory and experiment and slight deviation around H point. Notice also that the paramagnetic (1.125$\,T_c$) solution within the standard DFT has many unstable branches~\cite{PMphonon2}, which are here stabilized by proper description of the fluctuating moments existing above $T_c$.

\begin{figure}[!t]
\centering{
\includegraphics[width=1.0\linewidth,clip=]{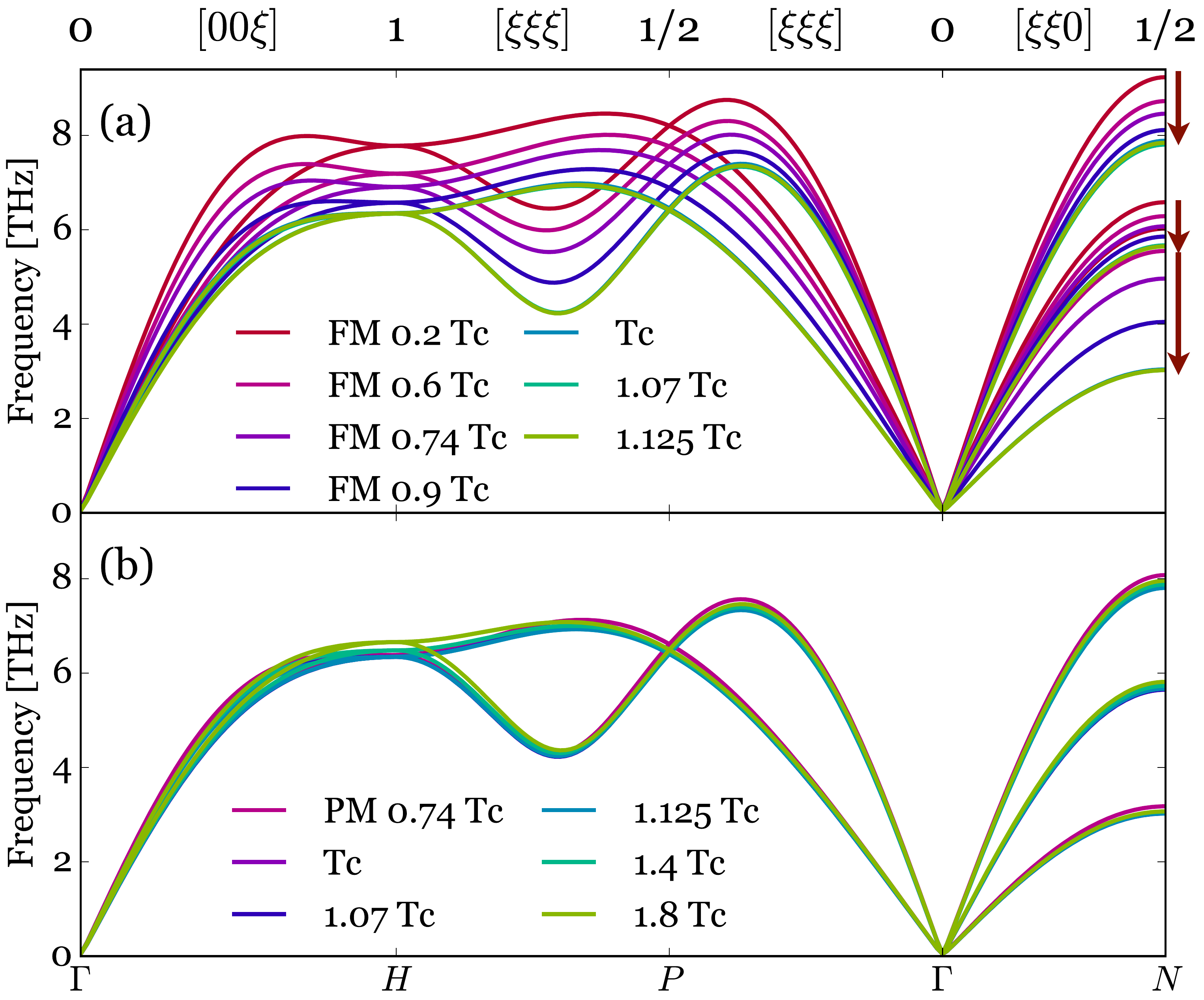}
}
\caption{(a) The calculated phonon dispersions in the BCC-$\alpha$ phase below and above Curie temperature at experimental equilibrium volume (b) phonons in the metastable paramagnetic BCC phase, but at constant volume (experimental volume at $T=T_c$).
}
\label{fig3}
\end{figure}
Next we show in Fig.~\ref{fig3}(a) the temperature dependence of the theoretically obtained phonon spectra in BCC phase from low temperature through magnetic transition and up to the $\alpha-\gamma$ transition. We notice very strong softening of the lowest branch at the $N$ point, which was shown to similarly soften experimentally in Refs.~\onlinecite{Phonon_softeningexp,AnharmonicsPhonon}, as well as substantial softening in the half-distance between $H$ and $P$ point. The arrows on the right mark the strong temperature variation of some phonon branches.
All these trends are very consistent with experiments. In Fig.~\ref{fig3}(b) we show phonon dispersion when the same calculation is done in metastable paramagnetic state below T$_c$, where experimentally only the ferromagnetic state is stable, and also far above T$_c$, in which FCC phase is thermodynamically more stable than the simulated BCC phase. In this paramagnetic calculation we fixed the volume to remove trivial quasi-harmonic effects on the phonon dispersion. We see that the phonon dispersion 
remains very similar up to very high temperature.

\begin{figure}[!t]
\centering{
\includegraphics[width=0.99\linewidth,clip=]{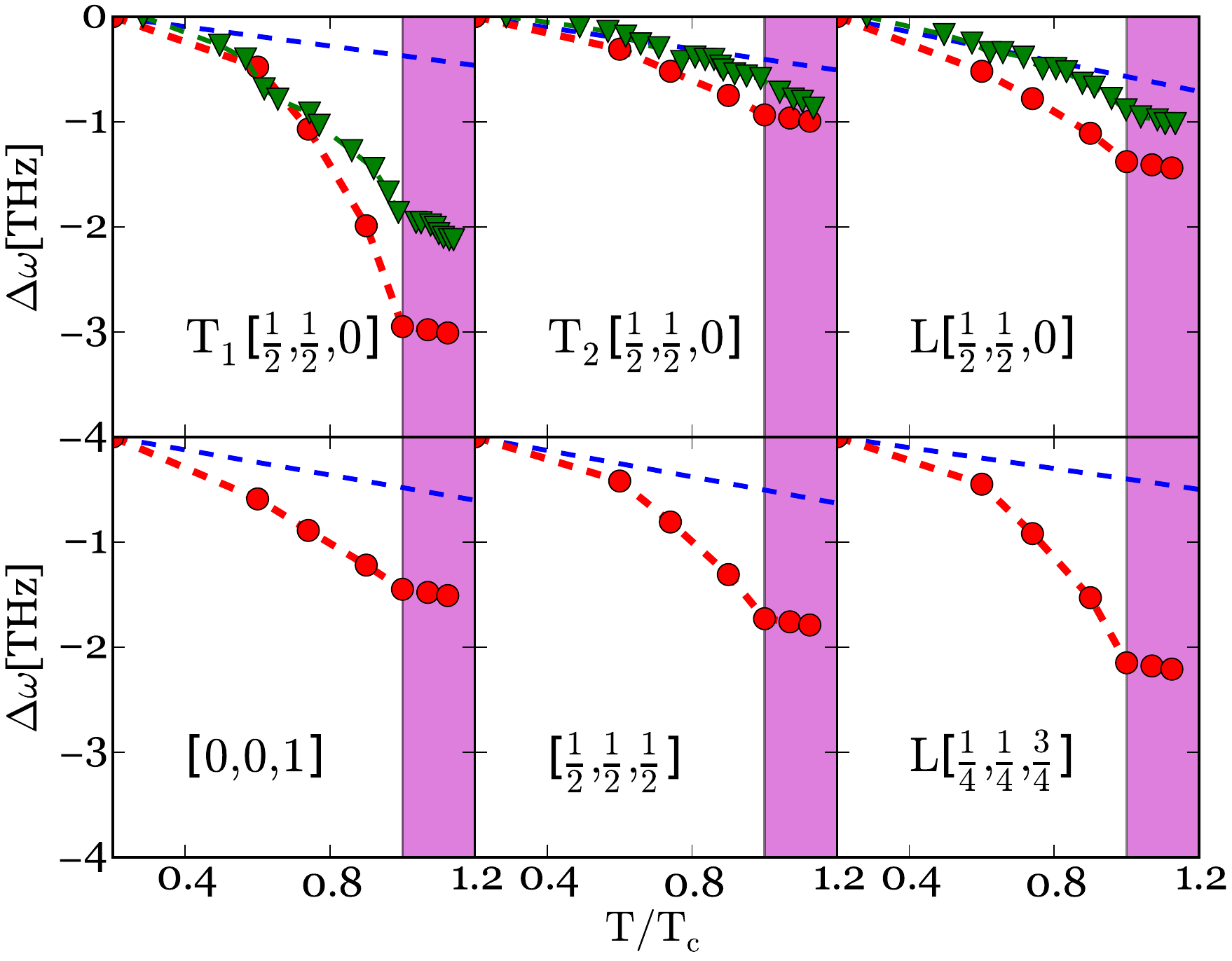}
}
\caption{ The change of the phonon frequencies for representative modes with temperature calculated by DFT+DMFT (red dots) compared to experimental data (green triangle). The dashed blue lines denote the change predicted by the quasiharmonic model: $\omega^{qh}(T)=\omega^{300K}(1-\gamma_{th})\frac{V_T-V_{300K}}{V_{300K}}$, where $\omega^{300K}$ is the calculated value of the phonon frequency at 300K, $V_T$ is the experimental volume of the unit cell at temperature $T$, and $\gamma_{th}$ is the thermal Gr\"uneisen parameter, approximated by a constant  
value of 1.81, as suggested in Ref.~\onlinecite{AnharmonicsPhonon,Grunratio}.}
\label{fig5}
\end{figure}
Fig.~\ref{fig5} shows the temperature dependence of selected phonon-branch frequencies and their comparison to quasi-harmonic approximation
(blue dashed line), which takes into account only the volume expansion. We notice inadequacy of such approximation, while the DMFT prediction, with melting of the long range order,  
is in reasonable agreement with experiment from Ref.~\onlinecite{AnharmonicsPhonon}. The experimental change is somewhat less abrupt at $T_c$ likely because the short range order persists  above $T_c$ in experiment.

On the basis of these results we can conclude that the phonon-softening, discovered experimentally many years ago~\cite{Expsoftening,Phonon_softeningexp}, is mainly due to melting of the magnetic long range order, and 
is not related to $\alpha\rightarrow\gamma$ phase transition, 
in contrast to what has often been assumed~\cite{Phonon_softeningexp}, and concluded in the previous DMFT study~\cite{PMphononnature}.
In our view, both the paramagnetic BCC and the FCC phase are dynamically stable at all temperatures, and their relative stability has to be determined by comparing their respective free energies.

\begin{figure}[!t]
\centering{
\includegraphics[width=0.99\linewidth,clip=]{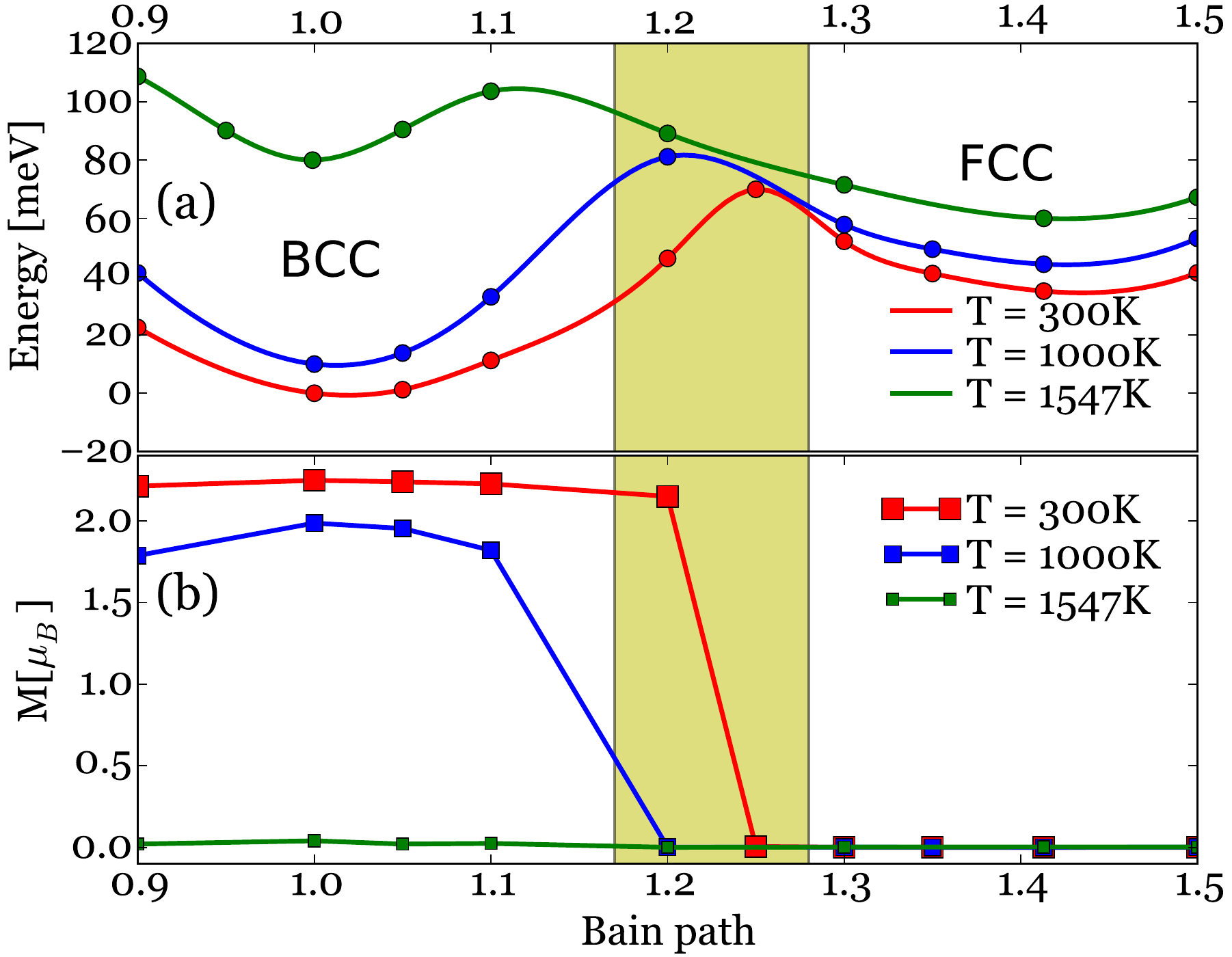}
}
\caption{ (a) The total energy computed along the Bain crystallographic transition from BCC to FCC phase in FM state. Note that at T=1547K the FM and PM phases are indistinguishable. (b) The ordered ferromagnetic moment along the same path.
}
\label{fig4}
\end{figure}

Since our results suggest that the phonon-softening mechanism in iron is unrelated to the $\alpha-\gamma$ structural transition, we want to demonstrate that our theory correctly predicts thermodynamic competition of the two phases (see related work of Ref.~\onlinecite{PRL_Leonov,Anisimov_2016}).  The martensitic $\alpha-\gamma$ transformations is as usually modeled by a continuous crystallographic transition from initial to the final phase, and in case of BCC-FCC transition the Bain path~\cite{Bain} is most often picked, which 
is described by a single parameter $c/a$ with $c/a=1$ corresponding to BCC and $c/a=\sqrt{2}$ to FCC phase. In Fig.~\ref{fig4} we show the total energy along this path, which clearly shows the double-well profile, characteristic of the first order phase transition, that does not require softening of phonons for the existence of the phase transition.
At low temperatures ($T=300\,$K, $1000\,$K), the global minimum is at the BCC structure ($c/a=1$) and at high temperature ($T=1547\,$K), it is at the FCC structure ($c/a=\sqrt{2}$). Along the path the ferromagnetic long range order disappears in our simulation, and at that value of $c/a$ (yellow region) the double-well curve reaches a maximum. At high temperature ($T=1547\,$K), where the ferromagnetic long range order disappears for all values of $c/a$, the total energy still keeps the double-well shape with very small total energy difference between BCC and FCC phase ($20\,$meV), in contrast to the DFT prediction, which  has a single minimum with both the magnetic and non-magnetic functionals.

The importance of disordered localized magnetic moments in paramagnetic phases of iron was stressed early on in the pioneering work of Grimvall~\cite{Grimvall}. This physics now emerges from a quantitative first principles method, and its implications for many physical quantities has been elucidated. We predict that the softening of the phonons in BCC structure is not related to its first order $\alpha$ to $\gamma$ transition, but it is due to the melting of the long range magnetic order. Our prediction can be checked by measuring the phonon dispersion of the paramagnetic iron under applied magnetic field, to check that long range magnetic order in the field hardens the phonons at selected points in the Brillouin zone. We predict that the BCC structure is dynamically stable at all temperatures, and is only thermodynamically unstable due to lower free energy of the FCC-$\gamma$ phase at the intermediate temperatures between the $\alpha$ and the $\delta$ phase.

We acknowledge the support of NSF DMR-1405303 (K.H.) and the Simons Foundation (Q.H.). T.B. was supported by the National Science Foundation (DMREF-1629260).
\bibliographystyle{apsrev4-1} 
\bibliography{iron} 

\newpage

%
%
%
%

\end{document}